\documentclass[conference]{IEEEtran}
\IEEEoverridecommandlockouts
\usepackage{cite}
\usepackage{amsmath,amssymb,amsfonts}
\usepackage{algorithmic}
\usepackage{graphicx}
\usepackage{textcomp}
\usepackage{xcolor}
\usepackage{csquotes}
\usepackage{subcaption}
\usepackage{dsfont}
\usepackage{cite}
\usepackage{balance}
\usepackage{textcomp}
\usepackage{algorithm}      
\usepackage{algorithmic}

\usepackage{varwidth}

\begin{document}


\title{Impact of Communication Loss on MPC based Cooperative Adaptive Cruise Control and Platooning}

\author{Mahdi Razzaghpour$^*$, Shahriar Shahram$^*$, Rodolfo Valiente$^*$, Yaser P. Fallah$^*$\\
$^*$Connected and Autonomous Vehicle REsearch Lab (CAVREL),\\
University of Central Florida, Orlando, FL\\
 razzaghpour.mahdi@knights.ucf.edu}



\maketitle

\begin{abstract}
Cooperative driving, enabled by communication between automated vehicle systems, is expected to significantly contribute to transportation safety and efficiency. Cooperative Adaptive Cruise Control (CACC) and platooning are two of the main cooperative driving applications that are currently under study. These applications offer significant improvements over current advanced driver assistant systems such as adaptive cruise control (ACC). The primary motivation of CACC and Platooning is to reduce traffic congestion and improve traffic flow, traffic throughput, and highway capacity. These applications need an efficient controller to consider the computational cost and ensure driving comfort and high responsiveness. The advantage of Model Predictive Control is that we can realize high control performance since all constrain for these applications can be explicitly dealt with through solving an optimization problem. These applications highly depend on information update and Communication reliability for their safety and stability purposes. In this paper, we propose a Model Predictive Control (MPC) based approach for CACC and platooning, and examine the impact of communication loss on the performance and robustness of the control scheme. The results show an improvement in response time and string stability, demonstrating the potential of cooperation to attenuate disturbances and improve traffic flow.
\end{abstract}



\section{Introduction}
Automated vehicles (AVs) depend only on the information collected by ego vehicle's sensors, i.e. camera, lidar, and radar. On the other hand cooperative systems use information augmentation and sensor fusion algorithms to combine ego vehicle's sensors data with information from Vehicle to everything (V2X) communication.

Adaptive Cruise Control (ACC) is a radar-based system, which is designed to enhance driving comfort and convenience by relieving the driver of the need to adjust the vehicle's speed to match the speed of the preceding vehicle. The information received from radar measures the distance to the preceding car in the same lane. Another capability of this sensor is to measure the relative velocity of the preceding vehicle in the same lane. ACC does not depend on inter-vehicular communication or any form of cooperation between vehicles. Therefore, ACC is not able to provide stable and robust performance with close spacing between vehicles. In recent years the research community works on two extensions of ACC called Cooperative Adaptive Cruise Control (CACC) and Platooning, which relies both on sensor measurements and wireless communication with other vehicles in the group. The difference between CACC and Platooning is the different gap regulation strategies. CACC uses a constant-time-gap-following policy whereas Platooning uses tightly coupled communication and control for constant clearance or constant distance gap. Typically, all members of a group broadcast their data like speed and acceleration to following vehicles in the group, allowing them to react faster to any changes in traffic conditions when compared to relying on on-board sensors only.

The majority of current works consider a string of 5 to 10 vehicles at most, while in this paper we increased the length of string to 25 vehicles and guaranty string stability. In addition, we use an All-predecessor-leader following (APLF) topology different from other Information Follow Topologies\cite{IFT}.

\begin{figure}[t]
\centering
\includegraphics[width=0.7\columnwidth]{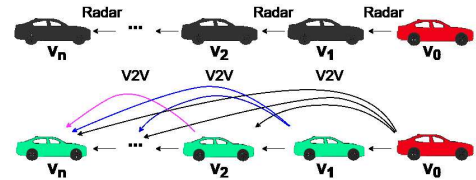}
\caption{A comparison between connectivity in ACC (top) and CACC or Platooning (bottom). In ACC every vehicle has access to information of the preceding one via radar. In CACC and Platooning cases, in addition to radar information, every vehicle has access to all preceding vehicles’ information via V2V.}
\label{fig:ACC_CACC}
\end{figure}

\section{Related Work}
Partners for Advanced Transit and Highways (PATH) project demonstrated a platoon of eight automated vehicles in 1997, following each other in close formation, with one of them changing lanes and shifting its position in the platoon formation \cite{path}. Joel \textit{et al.} shows that ACC can have only a small impact on highway capacity \cite{Effects_ACC}. On the other hand, Steven \textit{et al.} expresses that CACC has the potential to substantially increase highway capacity when it reaches a moderate to high market penetration \cite{Impacts_CACC}. Bart \textit{et al.} concluded that CACC can potentially double the capacity of a highway lane at high-CACC market penetration \cite{CACC_Char}.

The Grand Cooperative Driving Challenge (GCDC) is a project where multiple teams tested their CACC vehicles. GCDC 2016 aims at demonstrating how cooperative and automated vehicles can perform complicated platooning operations with close-to-reality traffic scenarios. The focus is the cooperative aspects, where advanced platoon operations (e.g., the cooperative merge of two parallel platoons) have been demonstrated. In addition, GCDC 2016 introduces a cooperative intersection scenario to demonstrate how cooperative vehicles solve complicated intersection scenarios safe and efficiently \cite{Grand_Challenge}.

The Energy ITS project in Japan aimed at the CO2 emission reduction from automobiles,  which includes two themes:  an implementation of automated truck platooning system and an evaluation method of effects of  ITS-related systems and technologies on the CO2 emission reduction\cite{Energy}.

Jeroen \textit{et al.} implemented CACC on a test fleet consisting of six-passenger vehicles adopting a constant time gap spacing policy \cite{Design}. Darbha \textit{et al.} had shown that the time headway can be minimized by using multiple predecessors’ information in the local controller \cite{Effects_V2V}. Studies in \cite{Heavy-Duty} have shown that a car with a velocity of 80km/h following only one predecessor at 25m achieves a 30\% reduction in aerodynamic drag, and a 40\% reduction can be attained by following two predecessors.

\begin{figure*}[h]
\centering
 \includegraphics[width=.8\textwidth]{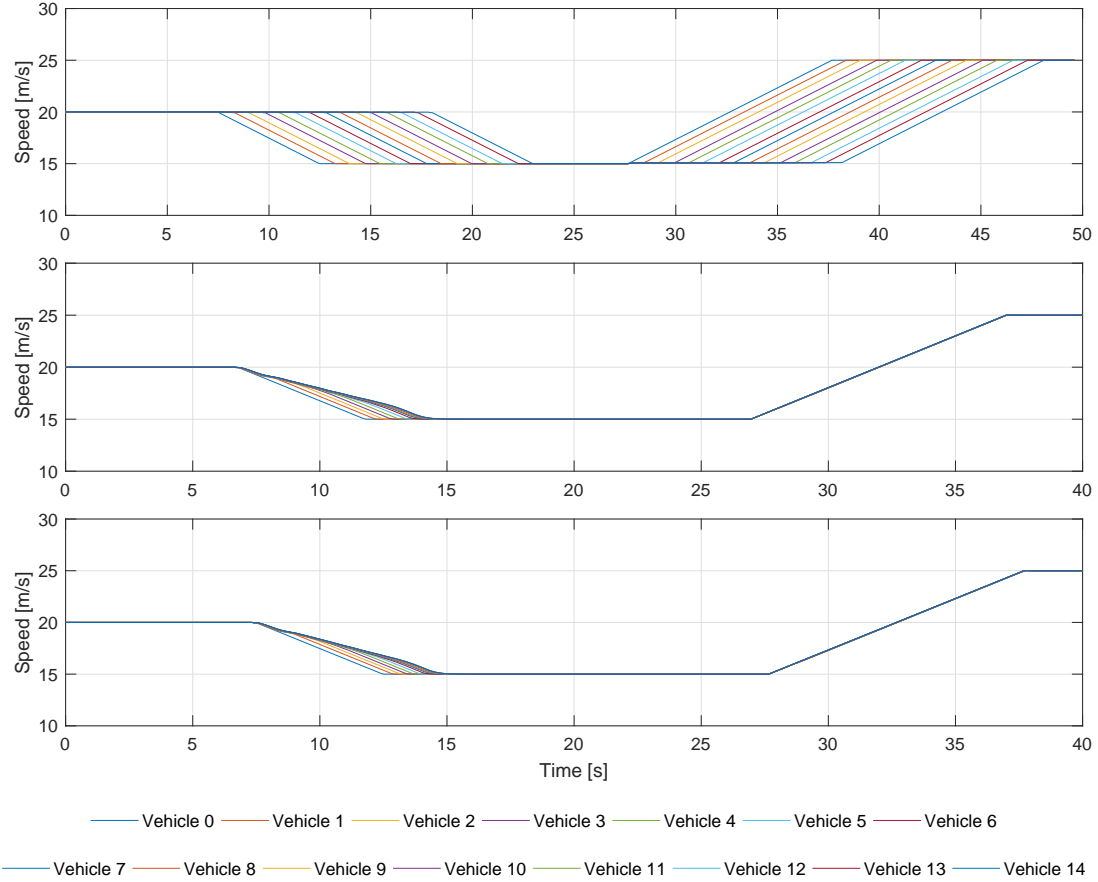}
    \caption{Speed profiles for Fifteen-vehicle test for ACC, CACC, and Platooning (from top to bottom)}
    \label{Figure:Speed}
\end{figure*}

\begin{figure*}[h]
\centering
 \includegraphics[width=.8\textwidth]{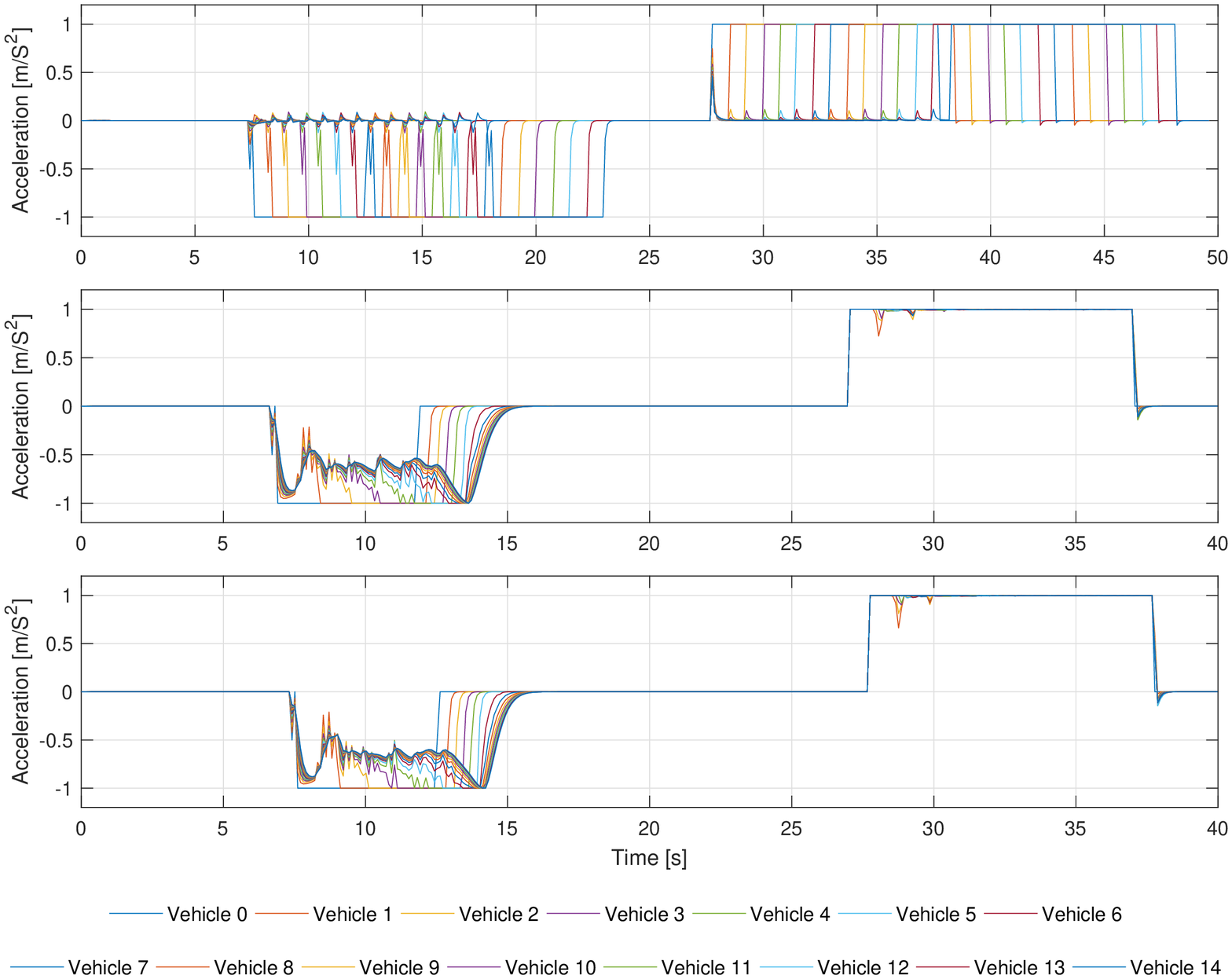}
    \caption{Acceleration profiles for Fifteen-vehicle test for ACC, CACC, and Platooning (from top to bottom)}
    \label{Figure:Acceleration}
\end{figure*}

\section{Background}
\subsection{ACC}
When there are multiple consecutive AVs in ACC, only after the second vehicle sensed, processed, and responded to the leader’s motion changes, the third vehicle can infer what happened to the first vehicle from the behavior of the second vehicle. In an autonomous ACC, the detection and response delay is cumulative from the leader to the downstream vehicles as shown in Fig  \ref{Figure:Speed}. The accumulated delay prevents string stability in long strings of the automated vehicles\cite{CACC_def}. Recent on-the-road experiments have shown that a stream of autonomous ACC vehicles is string unstable, resulting in a negative impact on lane capacity\cite{real_traffic}. By using Vehicle to Vehicle (V2V) Communication, every vehicle get information not only from its preceding, as the ACC case, but also from the vehicles in front of the preceding one as shown in Figure \ref{fig:ACC_CACC}.

Depending on the communication topology, different realizations of CACC and Platooning are possible. In this paper, we used All-predecessor-leader following (APLF) topology, and all the simulations are based on this topology. In APLF, in addition to radar information from the exact preceding vehicle, every vehicle has access to all preceding vehicles’ information and leader vehicle via V2V \cite{IFT}.

Each vehicle has a unique ID, i.e., $i = 0, 1, 2, \dots,N-1$, where i = 0 is the ID of the leader vehicle and the following vehicles have IDs $i \in\{1,2,\dots,N-1\}$. At each time slot t, all N vehicles broadcast their information to their following vehicles. By using both sensor and communication information, the vehicle string has a quicker response to any change in road traffic. The lead vehicle in the string (V0) is usually representing the surrounding traffic or the reference for the following vehicle. In our simulations, the lead vehicle (V0) is considered as a human-driven vehicle and provided with a reference trajectory to follow.

\subsection{Cooperation}
Using wireless communication, high-risk situations can be detected earlier via sharing situation awareness between vehicles and vehicles can follow their predecessors with higher accuracy, faster response, and shorter gaps to achieve system-wide benefits. Algorithms can use simple data from V2V communication e.g. speed, acceleration, location, or extensive forms of data like model-based communication\cite{model_based}, future intention\cite{GP,valiente2020dynamic}, performance limitation like maximum speed, acceleration, and braking capability for heterogeneous strings. For instance, intention sharing in these cases allows following vehicles to start to respond to any speed change, before the leader's speed changes measurably. Also via Infrastructure to Vehicle (I2V) Communication, vehicles can get long-term data like traffic ahead and speed recommendation.

CACC and Platooning highly depend on communication reliability and information updates, i.e. Basic Safety Massage (BSM) updates at 10 Hz rate, to maintain safety and stability. We used the same 10 Hz rate in our simulations. One upper limit that could be placed on string length is based on the range of the wireless V2V communication system, assuming that all CACC following vehicles will require direct communication from the lead vehicle in the string. High-performance communication technologies, Dedicated Short-Range Communications (DSRC)\cite{DSRC} and Cellular V2X (CV2X)\cite{CV2X} can provide at least 300 m of communication range to enable string lengths up to 15 to 25 vehicles. In our simulations, we consider strings of 5 to 25 vehicles. Also, we assumed that all vehicles are equipped with V2V communication.

\subsubsection{CACC}
In this system, gap policy is a constant time policy. In other words, the distance between vehicles is proportional to their speed, the higher the speed is, the longer the distance. CACC only provides longitudinal control of the vehicle's motion and the driver has to control the steering\cite{CACC_survey}.

Here, we consider a constant time headway spacing policy where the desired spacing is
\begin{equation}
D_{\text {desired i }}=D_{\text {safety }}+T_{\text {gap }} \cdot V_{i}
\end{equation}
where i is the vehicle index, $D_{\text {desired i }}$ is the desired gap between the ith vehicle and its predecessor, $D_{\text {safety }}$ is the minimum gap, $T_{\text {gap }} $ is the constant time headway representing the time that it will take the ith vehicle to arrive at the same position as its predecessor, and $V_{i}$ is the speed of the ith vehicle. Without loss of generality, in this paper, we assumed $D_{\text {safety }}$ is zero. Also, $T_{\text {gap }} $ in all simulations for CACC is $0.8$ second.

\subsubsection{Platooning}
Platooning has a very close coupling between vehicles to maximize highway capacity. In this architecture, the separation between vehicles remains constant and does not change as the vehicle speed changes. The primary goals of Platooning include increasing lane throughput and reducing aerodynamic drag by drafting which results in energy efficiency\cite{CPS_survey}. Platooning can improve fuel efficiency for vehicles and especially for trucks\cite{Energy}.

This discipline offers performance advantages, but the constant-clearance following is more difficult to achieve and requires a more formal platoon architecture. In addition interruptions in communication are more serious from a safety standpoint. In this paper, the desired distance for all Platooning simulations is $15$ m.






\begin{figure}[t]
\centering
\includegraphics[width=1\columnwidth]{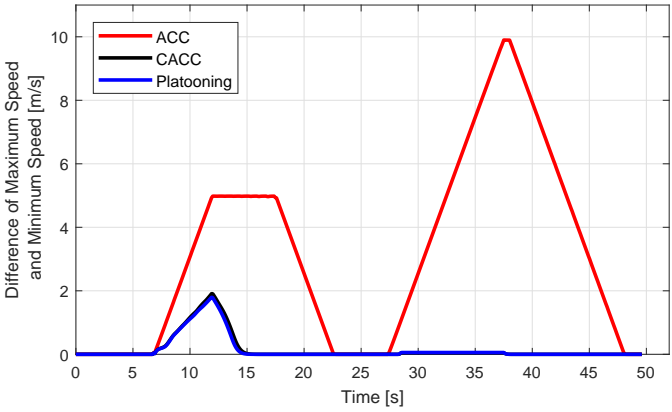}
\caption{Difference of Maximum speed and Minimum speed at each time step for the fifteen vehicle tests in Figure \ref{Figure:Speed}}
\label{Figure:diffSpeed}
\end{figure}

\begin{figure}[t]
\centering
\includegraphics[width=1\columnwidth]{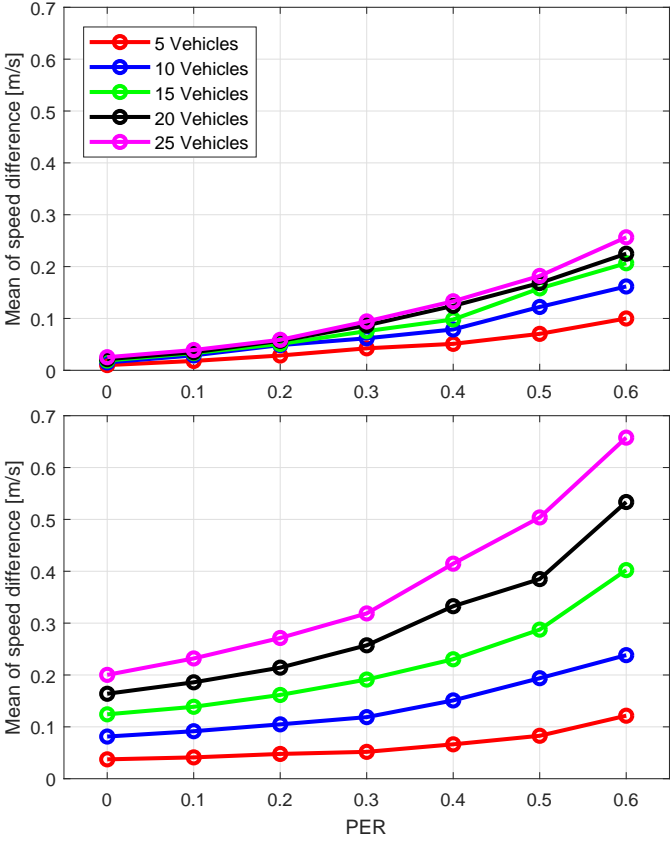}
\caption{Mean of speed difference for different string length Vs. different values of PER for CACC and Platooning (from top to bottom)}
\label{Figure:meandiffSpeed}
\end{figure}

\begin{figure}[t]
\centering
\includegraphics[width=1\columnwidth]{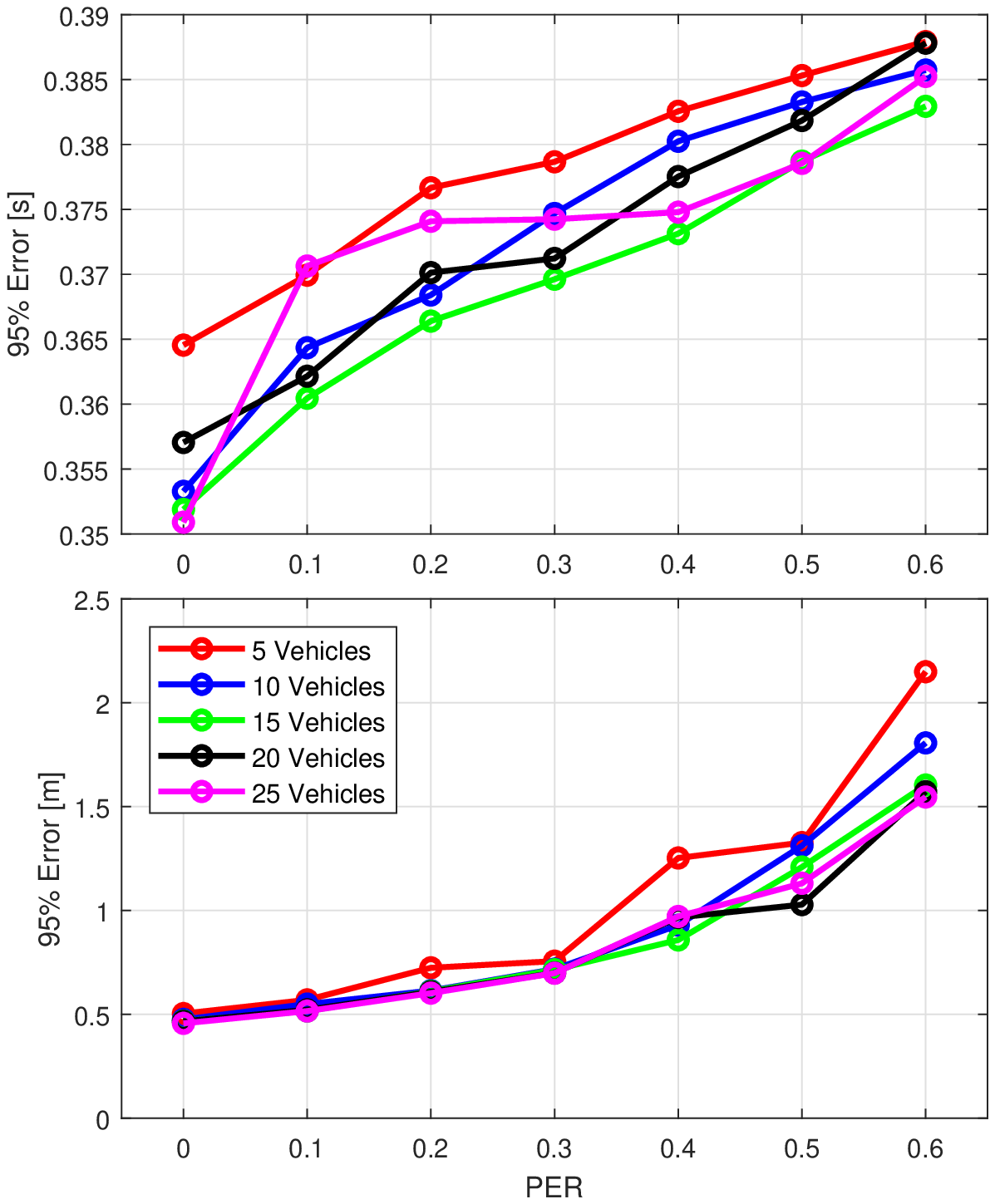}
\caption{95\% error for for different string length Vs. different values of PER for CACC and Platooning (from top to bottom)}
\label{Figure:95error}
\end{figure}

\section{Problem Statement}
The purpose of CACC and Platooning is to follow a desired inter-vehicle gap and the desired velocity between a group of preceding vehicles and a host vehicle. To validate these requirements, including minimizing the value of relative gap error and keeping the desired spacing policy, smoothing the acceleration and deceleration and providing a comfortable ride for passengers, keeping the velocity and acceleration at a reasonable interval, and providing the velocity adaptation mechanism in case of changing the velocity of the lead vehicle, we will introduce three kinds of control objective in this part.

\begin{itemize}
  \item \textbf{Collision avoidance:} It is the constraint with the highest priority and should never be violated under any circumstance. The goal of this constraint is to prevent the following vehicle from colliding with the preceding car. 
\begin{equation}
D_{i-1,i} \geq d_{\text {safety }}
\end{equation}
where $D_{i-1,i}$ is the distance between vehicles $i$ and $i-1$ 
  \item \textbf{Adaptive Speed Control:} As the objective of cooperation, all vehicles should be able to track the desired speed at each time instance.
  \begin{equation}
|v_{desired} - v_{i}|_{t} \longrightarrow 0
\end{equation}
where $i$ is the index of vehicles.

  \item \textbf{Driver Comfort:} To reduce the peak force and sudden jerk, acceleration should be bounded.
  \begin{equation}
|a_{i}|_{t} \leq Constant
\end{equation}
where $i$ is the index of vehicles.
\end{itemize}

Information from the radar is used to reduce gap error between the desired gap and the actual gap in ACC. In addition, for CACC and Platooning the local vehicle control objective can now be defined as regulating the error to be zero subject to the above-mentioned constraints for all preceding vehicles.

\subsection{Vehicle Model}
We define the state vector as:
\begin{equation}
Z = [x \hspace{2mm}  y \hspace{2mm}  \varphi \hspace{2mm}  v] 
\end{equation}
 Where $x$ is the x-position, $y$ is the y-position, \(\varphi\) is the yaw-angle and $v$ is the velocity. 
 The input vector is defined as:
\begin{equation}
u = [a \hspace{2mm}   \delta ] 
\end{equation}
Where $a$ is the acceleration and \(\delta\) is the steering angle.
  
Finally our vehicle model is:

\begin{equation}
\dot{Z}(t) = A'Z(t) + B'u(t)
\end{equation}
 
\begin{equation}
Z(t+1) = A Z(t) + B u(t) + C 
\end{equation}

Where $Z(t+1)$ is the output of control or new state and
 
\begin{equation}
A = 
\begin{pmatrix}
1 & 0 & cos(\bar{\phi})dt & - \bar{v}sin(\bar{\phi})dt \\
0 & 1 & sin(\bar{\phi})dt & \bar{v}cos(\bar{\phi})dt \\
0 & 0 & 1 & 0 \\
0 & 0 & \frac{tan(\bar{\delta})}{L}dt & 1
\end{pmatrix}
\end{equation}

\begin{equation}
B = 
\begin{pmatrix}
0 & 0 \\
0 & 0 \\
dt & 0 \\
0 & \frac{\bar{v}}{L {cos\textsuperscript{2}(\bar{\delta})}} dt
\end{pmatrix}
\end{equation}

\begin{equation}
C = 
\begin{pmatrix}
 \bar{v}sin(\bar{\phi}) \bar{\phi} dt \\
-\bar{v}cos(\bar{\phi})\bar{\phi}dt\\
0 \\
 \frac{\bar{v}\bar{\delta}}{L {cos\textsuperscript{2}(\bar{\delta})}} dt
\end{pmatrix}
\end{equation}

\begin{equation}
\dot{\phi} = \frac{v.tan(\delta)}{L}
\end{equation}

L is the length of vehicle (4.5 m) and the bar sign stands for the mean value of the variable between each simulation time.

\subsection{Model Predictive Control}
The advantage of Model Predictive Control (MPC) for our design is that it allows to incorporate operating constraints, i.e. regulates the velocity of the vehicle while satisfying the desired gap.

The control inputs are calculated by solving the General constrained optimization problem below during each simulation time step.

\begin{equation}
 \begin{aligned}
\begin{array}{c}
\min _{u} J=\sum_{k=1}^{T} (Z_{i, t+k}^{\top}Q_{i,i}Z_{i,t+k}+\Delta u_{i, t+k}^{\top}R_{\Delta u}\Delta u_{i,t+k}  \\
+u_{i, t+k}^{\top}R_{u}u_{i,t+k}) + \left(Z_{i,i-1}^{\top}  Q_{i, i-1} Z_{i, i-1}\right) \\
+\sum_{j=0}^{i-2}\left(Z_{i, j}^{\top}  Q_{i, j} Z_{i, j}\right) \\[10pt]

\text { subject to } \\[10pt]

Z_{t}^{min} \leqslant Z_{t+k} \leqslant Z_{t}^{max} \\[10pt]

u_{t}^{\operatorname{min}} \leqslant u_{t} \leqslant u_{t}^{max} \\[10pt]

\Delta u_{t}^{min} \leqslant \Delta u_{t} \leqslant \Delta u_{t}^{\operatorname{max}}
\end{array}
\end{aligned}
\end{equation}

where $t$ is the current time, $T$ is the prediction horizon, $\Delta u_{t}$ is the control input change, and the suffix “min” and “max” denotes lower bound and upper bound. $Q_{i, i}$, $R_{\Delta u}$, $R_{u}$ are the positive-semi definite weight matrices for the ego vehicle trajectory error, the change rate, and the magnitude of the control input, respectively. The second summation term is the place where MPC Consider ACC, and the third term is where MPC Consider CACC or Platooning based on the application. $Z_{i, j}$ is the relative measurement between vehicle $i$ and $j$ where relative gap and relative speed are considered. $Q_{i, j}$ is a matrix with weights for relative gap and relative speed. In this paper $Q_{i, j}$ is identical for different $i$ and $j$. The linear MPC optimization problem above results in a Quadratic Programming (QP) problem.

MPC needs to solve an optimization problem considering future prediction in every simulation period. So MPC requires a plant model to predict the future behavior of the plant along a finite time horizon. At each simulation instant, optimal control inputs are calculated for 10 prediction horizons. The control signal for the current instant is applied to the system and future states of the system are predicted by using the aforementioned vehicle model.

\subsection{Communication Loss}
In CAVs, vehicles frequently communicate their state information (e.g., position, speed, and heading) to other vehicles (e.g., every 100ms) over a broadcast wireless link. Also, each vehicle should keep track of the movements of neighboring vehicles based on the received information. These two sub-components of communication/networking and estimation together provide real-time situational awareness.

In this paper, we assume that the packet delivery succeeds with the same constant probability on all communication links. Packet loss is an independent and identically distributed random process\cite{Communication_Aspects}. At any given instance, the data is either received or lost. Each packet is assumed to be received with probability $P$ and missed with probability $1-P$ which is the same as PER \cite{Non-Ideal_Communication}.

For each vehicle $i$, if the packet from its neighbor $j$ is transmitted successfully, then $i$ will use the newly received information for local control. Otherwise, vehicle $i$ will update the latest received information based on Constant speed for the local control. In this scheme, we considered that the vehicle's speed will remain unchanged during each inter-packet gap. The algorithm uses the speed received from the last received information until it receives the next packet.

\subsection{String Stability}
The Goal of String Stability is to guarantee the reduction of disturbances propagated from the leading vehicle to the rest of vehicles in a string for a specific design, avoiding that leading vehicle speed changes cause amplification in the rest of the vehicles. It can be defined as 
\begin{equation}
|S S(s)|=\left|\frac{X_{i}(s)}{X_{i-1}(s)}\right| \leq 1, \quad i \geq 2
\end{equation}
where $i$ indicates the index of the vehicle in the string.

Sinan \textit{et al.} obtains bounds on Maximum Allowable Transmission Intervals (MATI) and Maximum Allowable Delays (MAD) while string stability is still guaranteed \cite{ss_cc}. For various initial conditions quantitative numbers for $T_{MATI}$ and $T_{MAD}$ can be obtained with the help of a numerical search algorithm. These times are around 1.5 seconds. In our Simulations, we use 10 Hz communication rate. As it is proved by results, our architecture is stable. Note that if the communication rate is sufficiently high, then the probability of receiving a packet will be large even if the PER is high.

The analysis shows that the ACC string is stable if there are relatively large headway-time values (larger than 0.5 sec), whereas the CACC system is string stable for all headway-time values that were considered. For additional analysis refer to \cite{IFT}.

\section{Results}

\begin{figure}[t]
\centering
\includegraphics[width=1\columnwidth]{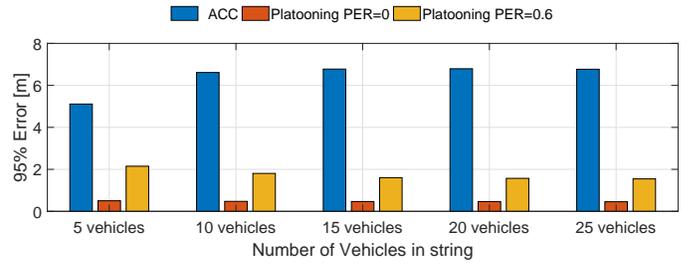}
\caption{95\% error comparison for different string length for 3 scenarios : ACC , Platooning best case scenario (PER=0) and Platooning worse case scenario (PER=0.6)}
\label{Figure:ACC_error}
\end{figure}

\begin{figure}[t]
\centering
\includegraphics[width=1\columnwidth]{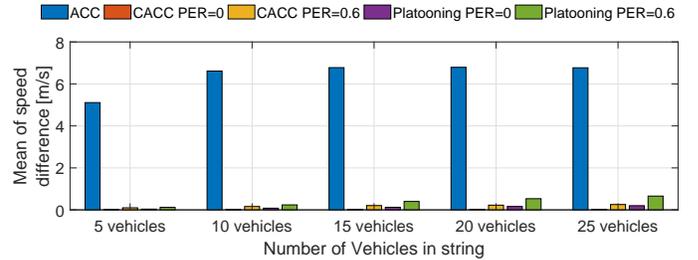}
\caption{Mean of speed difference comparison for different string length for 5 scenarios : ACC , CACC best case scenario (PER=0) and CACC worse case scenario (PER=0.6), Platooning best case scenario (PER=0) and Platooning worse case scenario (PER=0.6)}
\label{Figure:ACC_speed_var}
\end{figure}



 The speed, and acceleration response of a 15 vehicle string for ACC, CACC, and Platooning are shown in Figure \ref{Figure:Speed} and Figure \ref{Figure:Acceleration}. In all plots, the leader was supposed to reduce its speed from $20\ m/s$ to $15\ m/s$ and accelerate to reach to $25\ m/s$.  As Figure \ref{Figure:Speed} shows, in the case of ACC, any change in leader's speed propagated in the string with a delay as discussed in the Background. Each vehicle is controlled to follow its predecessor while maintaining a desired distance. These simulations involve 15 vehicles, with a maximum acceleration of $1 \ m/s^{2}$. It takes around 15 seconds for decelerating and around 20 seconds for accelerating that the whole string adapts to the new speed. CACC works based on a constant time gap, and Platooning works based on a constant distance gap. Using the same setup for CACC and Platooning, in both cases, it takes less than 10 seconds for decelerating and around 10 seconds for accelerating that the whole string adapts to the new speed. Also, it was seen that as we increase the index for vehicles (as we go away from the leader), the error value decreases. It is simply because those vehicles start to respond to changes very quickly.

For the rest of the simulations, we used 5 to 25  vehicles in the string and run multiple experiments for different PER values and the leader's trajectory has a time-varying speed with a maximum of $20 \ m/s$ and a minimum of $10 \ m/s$ to makes it look like a realistic highway driving scenario. To capture a statistical sense of worst-case behavior, and due to characteristics of wireless networks, we choose 95 percentile of the error's absolute value. As in \cite{CACC_def} we defined the error for CACC as the absolute value of the difference between actual time gap and desired time gap $(0.8 s)$ in seconds, while in platooning the error is defined as the absolute value of the difference between actual distance gap and desired distance gap $(15 m)$ in meters. 

The difference between Maximum speed and Minimum speed considering all string members at each time step is a good measure for traffic flow, we defined this metric as speed difference and used it to measure the performance. In Figure \ref{Figure:diffSpeed} we plot the speed difference for ACC, CACC, and Platooning for the results of the simulations presented in Figure \ref{Figure:Speed}.  

Figure \ref{Figure:meandiffSpeed} shows the mean of speed difference for CACC and Platooning for different string length and in variable PER scenarios. Platooning is more sensitive to packet loss compared to CACC. By increasing the length of the string, the average speed difference increase in both cases as expected. For 25 vehicles and PER of $0.6$ in platooning, it reaches to $0.67 \ m/s$ while in CACC it reaches to $0.26 \ m/s$. In Figure \ref{Figure:95error} we showed the 95\%  error for CACC and Platooning for various string length and different values of PER.
 
Figure \ref{Figure:ACC_error} shows the 95\%  error for Platooning and ACC. It shows that Platooning highly reduces the 95\%  error even at the highest PER (worse case scenario) compared to ACC. CACC error is measured in seconds, so we did not put it in this comparison. Finally Figure \ref{Figure:ACC_speed_var} shows that Platooning and CACC reduce the average speed oscillations and allow the traffic flow to be maintained at peak throughput even at the highest PER compared to ACC.





\section{Conclusion}
In this paper, we explored the performance of the CACC and Platooning with random loss. In contrast with ACC, CACC and Platooning showed improvements in response time and string stability, indicating the potential for a cooperative system to attenuate disturbances and improve traffic flow stability. We also provide statistics for errors in gap policy and speed variation of the string members for different channel situations (PER from 0 to 0.6) and different lengths of string (from 5 vehicles to 25 vehicles). Compared to the constant distance policy (Platooning), the constant time headway spacing policy (CACC) improves scalability and string stability. Even though a larger time headway can improve the string stability in case of random loss, but it also negatively affects some of CACC benefits such as increasing road throughput. The study shows that communication rate and loss ratio have a significant influence on the string stability and performance of the proposed architecture. Reducing speed oscillations will reduce vehicle crashes and improve traffic flow and comfort.
In the future, we will model a realistic link model for highway scenarios, and use that model as a source of uncertainty and randomness. The effectiveness of the proposed control algorithm is carried out only numerically and additional experimental study is needed in the future to support the numerical results.

\section{Acknowledgement}
This research was supported by the National Science Foundation under grant number CNS-1932037.

\balance
\bibliography{refs.bib}{}
\bibliographystyle{unsrt}
\end{document}